\documentclass[preprint,aps,amsmath,amssymb,superscriptaddress]{revtex4}
\usepackage[cp1250]{inputenc}

\begin{document}

\title{On plane gravitational waves in real connection variables
}

\author{Franz Hinterleitner}
\affiliation{Department of Theoretical Physics and Astrophysics,
Faculty of Science of the Masaryk University, Kotl\'{a}\v{r}sk\'{a}
2, 611\,37 Brno, Czech Republic}
\author{Seth  Major}
\affiliation{Department of Physics, Hamilton College, Clinton NY
13323
USA}

\date{December 2010}

\begin{abstract}
We investigate using plane fronted gravitational wave space-times as
model systems to study loop quantization techniques and dispersion
relations.  In this classical analysis we start with planar
symmetric space-times in the real connection formulation. We reduce
via Dirac constraint analysis to a final form with one canonical
pair and one constraint, equivalent to the metric and Einstein
equations of plane-fronted-with-parallel-rays (pp) waves. Due to the
symmetries and use of special coordinates general covariance is
broken. However, this allows us to simply express the constraints of
the consistent system. A recursive construction of Dirac brackets
results in non-local brackets, analogous to those of self-dual
fields, for the triad variables. Not surprisingly, this classical
analysis produces no evidence for dispersion, i.\,e. a
variable propagation speed of gravitational pp-waves.
\end{abstract}

\maketitle PACS 04.20.Fy, 04.30.Nk, 04.60.Ds, 04.60.Pp

\section{Introduction}

The success in constraining modified dispersion relations
\cite{jlm,limits,lm_review,jlm_rev} has renewed efforts to see
whether, in the context of various approaches to quantum gravity,
such modifications arise.   This is interesting even in model
systems where the quantization may be more unambiguously carried out
and where it is possible to identify the origin of the
modifications, should they appear. For instance, this has been
explored in the context of polymer quantization of scalar fields in
flat space-time \cite{hhs,hk}. In this case the origin of the
modifications lies in the choice of classical polymer variables, in
particular the length scale required to express the exponentiated
momentum variable, rather than in a granularity of spatial geometry.
While loop quantum gravity there are heuristic
results suggesting that there might be modifications to dispersion
relations \cite{GP,AM-TUneut,AM-TUphot}, it would be interesting to
investigate possible modifications in a model system in which both
the origin of the modification is clear and in which the
quantization may be completed. This paper explores whether the
symmetry reduced space-times of
plane-fronted gravitational waves with parallel rays may be a
suitable context in which to explore modifications to dispersion
relations.

Classical plane gravitational waves are, like homogeneous and
isotropic
cosmological models, among the most simple exact solutions of
General Relativity (GR). Despite the nonline\-arity of GR, due to the
symmetry of  the model
pulses of plane fronted with parallel rays (pp) waves travel without
dispersion and leave space flat outside the pulse; they form a
``wave sandwich" with the gravitational wave pulse between regions
of flat space.  In fact the only gravitational waves with this flat
space sandwich property are pp-waves \cite{yurtsever}.  Due to their
simplicity pp-waves promise to be good candidates to test
quantization techniques for pure gravity.  It is also intriguing
that the Einstein equations for plane electromagnetic waves coupled
to gravity take the same form as the pp-waves \cite{MTW}, suggesting
that a quantization and study of dispersion relations of pp-waves
could be extended to this Einstein-Maxwell theory model. So the
quantization of pp-waves may yield an answer to whether quantum
effects of 3-dimensional geometry can lead to dispersion of
gravitational (or electromagnetic) plane waves, whether spatial granularity
can yield an energy-dependence of the speed of gravitation waves or light.

In view of this eventual goal, we formulate polarized parallel plane
waves in terms of the real connection variables, proceeding from
rather general assumptions about homogeneity in two dimensions, to a
form equivalent to a standard form in the literature, given below.
Despite the simplicity of this well-known result, the canonical way
to this goal is not trivial.

The metric of pp-waves propagating in $z$ direction,
in a ``Rosen-type" chart, given by Misner, Thorne, and Wheeler
\cite{MTW} is
\begin{equation}\label{metric}
{\rm d}s^2=- {\rm d}t^2 + L^2e^{2\beta}{\rm d}x^2 +
L^2e^{-2\beta}{\rm d}y^2 + {\rm d}z^2.
\end{equation}
(See also Ehlers and Kundt \cite{EK}.) This metric has a convenient
interpretation for our purposes. The function $L$, called the
``background factor", is determined by the free function $\beta$,
called the ``wave factor". Both $L$ and $\beta$ are functions of
$v:=t+z$ or $u:=t-z$. $L$ satisfies the Einstein equation
\begin{equation}\label{1stein}
L''+(\beta')^2 L=0.
\end{equation}
(In this equation the prime denotes a derivative with respect to
$u$ or $v$.) This single equation survives the reduction of GR.  In
it $\beta'$ acts as a ``time-dependent" angular frequency.  In the
light cone coordinates $u$ -- once $\beta(u)$, $L(0)$ and $L'(0)$ are
specified -- the function $L(u)$ is determined just as in a simple
1-dimensional mechanical system.

The above form of the metric as well as equation (\ref{1stein}) are
valid only for waves traveling in the positive {\em or} in the
negative $z$ direction. The combination of both, i.~e. colliding
waves, are a problem of a much higher degree of difficulty \cite{Gr,
GPod}.

An important feature of the above coordinate system is that it does
not globally cover the space-time inhabited by plane waves. One may
see this even with a short pulse, with $\beta\neq0$ only in an
interval $(u_-,u_+)$. For example choosing $L\equiv1$ in the region
of flat space in front of the pulse, at the location of the pulse,
where $\beta\neq0$, $L''$ becomes negative due to (\ref{1stein}) and
so $L$ decreases inside the wave, as long as $L\geq 0$. (We assume
the pulse to be short enough and not too strong so that $L>0$
everywhere inside.  For details of this approximation see
\cite{MTW}.)  In the flat space behind the pulse, when $\beta$ is constant
again, $L''=0$ and $L$ is a linear function, which has to be matched
smoothly to $L$ at the end of the pulse. This leads necessarily to
$L=0$ at a certain value $u_2>u_+$ somewhere behind the pulse, a
coordinate singularity in flat space \cite{Bondi,MTW}. In this case
the coordinate system is valid in the region $u<u_2$.

One can choose -- and we shall do so -- $L$ to be a non-constant
linear function on both sides of the wave pulse, with one zero in
front, at $u=u_1$, $u_1<u_-$, and one behind it. In this case, the
coordinates cover a slice of the gravitational wave sandwich.
In detail, we have the three sub-intervals of the coordinate range
$(u_1,u_2)$
\begin{quote}
(1) $u_1<u\leq u_-$, $L'=\mbox{const.}>0$, $\beta=0$, flat space in
front of the pulse,\\
(2) $u_-<u<u_+$, $L''<0$, $\beta\neq0$, the pulse, and \\
(3) $u_+<u<u_2$, $L'=\mbox{const.}<0$, $\beta$ constant, flat space behind
the pulse.
\end{quote}
The boundary conditions at $u=u_{1,2}$ are flat-space boundary
conditions with constant $\beta$ and $L=0$ at the coordinate singularities.
Despite the regions of flat space before and after the pulse,
neighboring test particles in the $xy$ plane accelerate and fall
towards each other as the wave passes \cite{Bondi,MTW}.

In the next section we review the canonical variables and the
polarized Gowdy model of \cite{BD1}.  The consistent reduction to
the pp-wave case is accomplished in sections \ref{Breduction} and
\ref{consistency}. The Dirac brackets are constructed in section
\ref{DB}.  Time evolution in the preferred coordinates is discussed
in section \ref{evolution}.  The results of the classical
calculations are summarized in section \ref{sum}. Finally, a note on
the orthogonality of the connection and the Immirzi parameter is in
\ref{orthog}.

\section{Symmetry Reduction}
\label{sym_reduce}

\subsection{The connection variables}
We formulate the system, after a 3+1 split, with the usual
densitized triads ${E^a}_i$ and connection components ${A_a}^i$ of
the real connection formulation. (See Ref. \cite{al_rev} for a review.)  We denote $a=x,y,z$
as a spatial and $i=1,2,3$ as an su(2) Lie algebra index. By
homogeneity, we further assume that on every spatial slice they are
functions of $z$ alone, with their time dependence to be determined
by equations of motion. The system has a close formal analogy to the
polarized Gowdy $T^3$ model analyzed by Banerjee and Date
\cite{BD1,BD2}. For this reason we have chosen essentially the same
notation, so that at many points we can refer to these papers. Of
course, the angular variable $\theta$ of \cite{BD1} had to be
changed to $z$ and, as is clear from this notation, we are also not
working with a compact spatial topology. 

The symmetry reduction from full GR to the Gowdy model is carried
out in a process outlined, for example, in the appendix of
\cite{LQC}. In the present case the symmetry reduction is briefly
the following: We start with a principal fibre bundle of $SU(2)$,
the gauge group of loop quantum gravity (LQG), over a
three-dimensional space manifold. The spatial symmetry in the
presence of pp-waves consists of translations in the $x$ and $y$
direction, the orbits of which are planes parallel to the
$xy$-plane. The space manifold is decomposed into an orbit bundle
with a one-dimensional basis manifold, the $z$-axis, called the
reduced manifold. The ``reduced bundle" is the trivial principal
$SU(2)$ bundle over a single coordinate neighborhood $z_1<z<z_2$ of
the reduced manifold, where $z_1$ and $z_2$ correspond to the null
coordinate boundaries $u_1$ and $u_2$ in the introduction. The
symmetry reduction proceeds with a decomposition of the bundle
connection. The latter one is separated into the reduced connection,
i.~e. the restriction along $z$, ${A_z}^i(z)\tau_i$, and into scalar
fields on the reduced manifold, ${A_x}^i(z)\tau_i$,
${A_y}^i(z)\tau_i$ where $\tau_i$ are the usual su(2) generators. By
a choice of gauge the reduced connection is assumed to lie in the
subalgebra generated by $\tau_3$, the scalar fields in the subspace
spanned by $\tau_1$ and $\tau_2$, so that the matrices ${A_a}^i$ and
the canonically conjugate densitized triad matrices become
block-diagonal \cite{martinspherical}
\begin{equation}
{E^z}_I=E^\rho_3=0, \hspace{6mm} {A_z}^I={A_\rho}^3=0,
\hspace{3mm} \text{ with }\rho=x,y,\hspace{3mm} I=1,2.
\end{equation}
As usual in LQG, the connection ${A_a}^i$ is defined as the
combination
\begin{equation}
{A_a}^i={\Gamma_a}^i-\gamma\,{K_a}^i
\end{equation}
of the torsion-free spin connection ${\Gamma_a}^i$ and the extrinsic
curvature ${K_a}^i$.  The Barbero-Immirzi parameter is denoted with
$\gamma$.
These variables are subject to the usual constraints
of canonical GR, the Gau\ss, the diffeomorphism, and the Hamiltonian
constraint.

Due to the planar symmetry of the waves, the phase space variables
are free of dependence on $x$ or $y$. We restrict integration in the
$xy$ plane to a finite, fiducial patch with area $A_o$. Integrating
the symplectic structure over this patch gives
\begin{equation}
\Omega = \frac{A_o}{\kappa' \gamma} \int {\rm d} z \left(  {\rm
d}A_z^3 \wedge {\rm d}E^z_3 + {\rm d}A_\rho^I \wedge {\rm d}E^\rho_I
\right)
\end{equation}
where $\kappa' = 4 \pi G$. For the rest of this article we use
$\kappa = \kappa'/A_o$.
Following \cite{BD1} we denote ${E^z}_3$ by $\cal E$ and ${A_z}^3$
by $\gamma \cal A$ and introduce polar coordinates in the ``1-2"
plane,
\begin{equation}
\begin{array}{ll}
{E^x}_1=E^x\cos\beta,\hspace{3mm} &{E^x}_2=E^x\sin\beta, \hspace{3mm}
\\
{E^y}_1=-E^y\sin\bar\beta, \hspace{3mm} &{E^y}_2=E^y\cos\bar\beta,
\end{array}
\end{equation}
\begin{equation}
\begin{array}{ll}
{A_x}^1=A_x\cos(\alpha+\beta), \hspace{5mm} &
{A_x}^2=A_x\sin(\alpha+\beta), \\[3mm]
{A_y}^1=-A_y\sin(\bar\alpha+\bar\beta), &
{A_y}^2=A_y\cos(\bar\alpha+\bar\beta).
\end{array}
\end{equation}
The canonically conjugate connection variables to the radial
variables $E^x$ and $E^y$, $\beta$ and $\bar\beta$ are
\begin{equation}
K_x:=\frac{1}{\gamma}A_x\cos(\alpha)\hspace{5mm}\mbox{and}\hspace{5mm}
K_y:=\frac{1}{\gamma}A_y\cos(\bar\alpha)
\end{equation}
and
\begin{equation}
P^\beta:=-E^x\,A_x\sin(\alpha)\hspace{5mm}\mbox{and}\hspace{5mm}
P^{\bar\beta}:=-E^y\,A_y\sin(\bar\alpha),
\end{equation}
respectively.

\subsection{The polarization condition}

Following Banerjee and Date \cite{BD1} we carry out a reduction
to the polarized model by setting $\beta=\bar\beta$.  This ensures
that the Killing vectors $\partial_x$ and $\partial_y$ are
orthogonal. This means that
$E^x$ and $E^y$ become orthogonal in the sense that
\begin{equation}\label{ort}{E^x}_i{E^y}_i=0
\end{equation}
and that the spatial part of the metric (\ref{metric}) becomes
diagonal.  Denoting the spatial distance as ${\rm d}s^2$
\begin{equation}
\label{met}
{\rm d}s^2={\cal E}\frac{E^y}{E^x}\,{\rm d}x^2+{\cal
E}\frac{E^x}{E^y}\,{\rm d}y^2+\frac{E^xE^y}{\cal E}\,{\rm d}z^2.
\end{equation}
At this point we do not yet specify the lapse function and the
shift vector. After redefinition of the angular variables and their
momenta
\begin{equation}
\xi:=\beta-\bar\beta,
\hspace{2cm}P^\xi:=\frac{P^\beta-P^{\bar\beta}}{2}
\end{equation}
and
\begin{equation} \eta:=\beta+\bar\beta,
\hspace{2cm}P^\eta:=\frac{P^\beta+P^{\bar\beta}}{2}
\end{equation}
the condition $\beta=\bar\beta$ may be imposed in the form of a
second-class constraint, $\xi:=\beta-\bar\beta=0$, whose Poisson
bracket with the Hamiltonian constraint does not vanish weakly. When
$\{\xi,H\}$ is added to the constraints, then together with $\xi$ it
forms a pair of second-class constraints, weakly Poisson-commuting
with all the other constraints. After introducing Dirac brackets,
this pair of constraints can be imposed strongly, thus eliminating
$\xi$ and $P^\xi$ \cite{BD1}.

The remaining phase space variables are ${\cal A}$, $K_x$, $K_y$, $\eta$ and
$\cal E$, $E^x$, $E^y$, $P^\eta$; their Dirac brackets are equal to
the Poisson brackets, as they all have weakly vanishing Poisson
brackets with the two strongly imposed constraints.

\subsection{The gauge constraint}

We have accomplished the symmetry reduction and imposed the
polarization condition. In these variables the Gau\ss\ constraint
$G$, the diffeomorphism constraint $C$ and the Hamiltonian constraint
$H$ reduce to (see \cite{BD1})
\begin{eqnarray}
&& G=\frac{1}{\kappa\gamma}\,[{\cal E}'+2P^\eta],\\
&& \label{diffeo}C=\frac{1}{\kappa}\,[K_x'E^x+K_y'E^y-{\cal E}'{\cal
A}+\frac{1}{\gamma}\,\eta'P^\eta],\\ && H=
-\frac{1}{2\kappa\sqrt{E}}\left[\frac{\kappa^2}{2}\,G^2+
(K_xE^x+K_yE^y)\left(\frac{\eta'}{\gamma}+2{\cal A}\right){\cal
E}+2K_xE^xK_yE^y+\right.\\
&& \hspace{8mm}\left.\displaystyle\frac{1}{2}\left\{({\cal
E}')^2-{\cal
E}^2\left(\displaystyle\frac{{E^y}'}{E^y}-\frac{{E^x}'}{E^x}\right)^2\right\}+
2\left\{-\left(\displaystyle\frac{{E^x}'}{E^x}+\frac{{E^y}'}{E^y}\right)
{\cal E}P^\eta+{\cal E}'P^\eta+2{\cal
E}{P^\eta}'\right\}\right]\nonumber,
\end{eqnarray}
where $E:={\cal E}E^xE^y$ is the determinant of the 3-metric. The
prime means derivative with respect to $z$ and $\kappa$ is the
gravitational constant. The total Hamiltonian is
\begin{equation}
H_{\text{tot}} = A_o \int dz \left( \lambda G + n C + N H \right).
\end{equation}
The authors of \cite{BD1} point out that for the polarized Gowdy
model an orthogonality condition on the ${A_a}^i$ analogous to
(\ref{ort}), namely
\begin{equation}\label{aort}
{A_x}^i{A_y}^i=\gamma\left(K_x{E^x}'-K_y{E^y}'\right)=0
\end{equation}
in the above variables, is not conserved under evolution. More
precisely, the Poisson bracket of the condition with $H$ does not
vanish weakly, it would give rise to a further constraint, and so
on, rendering the system inconsistent. Nevertheless, a careful
analysis of the spin connection and extrinsic curvature derived from
the metric (\ref{metric}) shows that pp-waves do satisfy this
condition. We will come back to this issue in Section \ref{orthog}.

For the present we follow the simplifications in \cite{BD1} one more
step by strongly imposing the Gau\ss\ constraint together with the
associated gauge fixing condition $\eta=0$ and thus remove the
variables $\eta$ and $P^\eta$. Again the Dirac brackets of the
remaining variables are equal to their Poisson brackets, defined by
\begin{eqnarray}
\label{poisson}
&&\{F,G\}=\kappa \int{\rm d}z\left(\frac{\delta F}{\delta \cal
A}\frac{\delta G}{\delta\cal E}-\frac{\delta F}{\delta\cal
E}\frac{\delta G}{\delta\cal A}+\frac{\delta F}{\delta
K_\rho}\frac{\delta G}{\delta E^\rho}-\frac{\delta F}{\delta
E^\rho}\frac{\delta G}{\delta K_\rho}\right),
\end{eqnarray}
When we have carried out this, the diffeomorphism
constraint (\ref{diffeo}) drops its last term and the Hamiltonian
constraint becomes equal to
\begin{equation}\label{ham}\begin{array}{ll}
H=-\displaystyle\frac{1}{2\kappa}\frac{1}{\sqrt{E}}\left[
2K_xE^xK_yE^y+2(K_xE^x+K_yE^y){\cal E}{\cal A}\right]+\\[4mm]
\displaystyle\frac{1}{4\kappa}\frac{1}{\sqrt{E}}\left[{\cal
E}^2\left(\displaystyle\frac{{E^y}'}{E^y}-\frac{{E^x}'}{E^x}\right)^2-
2{\cal
EE}'\left(\displaystyle\frac{{E^y}'}{E^y}+\frac{{E^x}'}{E^x}\right)
+({\cal E}')^2+4{\cal EE}''\right].
\end{array}\end{equation}
This last step reduced the system to three canonical pairs, related
by two first-class constraints, so that the field theory has exactly
one phase space degree of freedom per spatial point. As shown in Ref. \cite{BD1} the
algebra of the constraints is the correct one for canonical GR.

\subsection{Reduction to pp-waves and the spatial Einstein equation}
\label{Breduction}

Now we reduce the theory to a model equivalent to the one formulated
in metric variables by (\ref{metric}) and (\ref{1stein}), again
using Dirac's constraint analysis. The metric of (\ref{metric})
contains two functions, $L$ and $\beta$ rather than the three
functions $\cal E$, $E^\rho$ in (\ref{met}). The one Einstein
equation (\ref{1stein}) is not equivalent to the remaining
constraints $C=0$ or $H=0$. So we need (at least) one more
constraint. Comparing the spatial part of the metrics (\ref{metric})
with (\ref{met}) we see that we need the primary constraint
$g_{zz}=1$, or
\begin{equation}
B:={\cal E}-E^xE^y=0.
\end{equation}
Of course there is no guarantee that the resulting system is
consistent -- after all the polarized Gowdy is already reduced to
two phase space degrees of freedom -- but this system is simple
enough so we can introduce the appropriate constraints in the
special coordinate system.

For the reduced theory to be consistent, the new local constraint
$B$ must be  preserved under evolution of the total Hamiltonian
constraint. The Poisson bracket with the smeared-out diffeomorphism
constraint
\begin{equation}
C[n]:=A_o \int{\rm d}z\,n(z)C(z)
\end{equation}
is
\begin{equation}
\{B(z),C[n]\}=-n(z)B'(z)+2n'(z)E^x(z)E^y(z).
\end{equation}
Generally this is not weakly equal to zero, because the new
constraint is not invariant under local diffeomorphisms due to the
different nature of $\cal E$ and $E^\rho$. The variable $\cal E$
transforms as a scalar, whereas $E^x$ and $E^y$ transform as scalar
densities, as can be seen from the Poisson brackets
$$
\{{\cal E}(z),C[n]\}=n(z){\cal E}'(z), \hspace{5mm}
\{E^\rho(z),C[n]\}=(n(z)E^\rho(z))'.
$$
Only when the test function $n$ is constant along $z$ -- meaning the
shift vector depends only on $t$  -- is the combination $B$ of $\cal
E$ and $E^\rho$ meaningful and conserved under the action of $C[n]$.

In the end the failure of $B$ to be diffeomorphism invariant is not
a surprise. Demanding $g_{zz}=1$ we obviously restrict local
diffeomorphism invariance. (In the special case (\ref{metric}) the
shift vector is equal to zero.) The local constraint $C(z)$, in
contrast to the global translation generator $C[n]$, becomes
second-class after introducing the new constraint $B(z)$.

With the Hamiltonian constraint, smeared out with a (lapse) function
$N$, the constraint $B$ has the Poisson bracket
\begin{equation}
\{B(z),H[N]\}=N\left(\frac{(K_xE^x+K_yE^y)}{\sqrt{E}}\,B-2\sqrt{E}\cal
A\right).
\end{equation}
$B$ will thus be preserved under the evolution generated by $H$,
only if we add a new constraint
\begin{equation}
{\cal A}=0.
\end{equation}
The constraints $\cal A$ and $B$ form a second-class conjugate pair
\begin{equation}
\label{pair}
\{{\cal A}(z),B(z')\}=\kappa \delta(z-z').
\end{equation}

The new constraint $\cal A$ must be preserved as well. It is
diffeomorphism invariant in the full sense, and thus translation
invariant, since
\begin{equation}
\{{\cal A}(z),C[n]\}=[n(z){\cal A}(z)]'
\end{equation}
is weakly equal to zero.
The Poisson bracket of $\cal A$ with the Hamiltonian constraint is
\begin{equation}\label{AH}\begin{array}{ll}
\{{\cal A},H[N]\}=\!\!&\!\kappa N\left[\displaystyle\frac{\partial
H}{\partial{\cal E}}-\left(\frac{\partial H}{\partial{\cal
E}'}\right)'+\left(\frac{\partial H}{\partial{\cal
E}''}\right)''\right]\\[4mm]
&-\kappa N'\left[\displaystyle\frac{\partial H}{\partial{\cal
E}'}-\left(\frac{\partial H}{\partial{\cal
E}''}\right)'\right]+\kappa N''\displaystyle\frac{\partial
H}{\partial{\cal E}''} . \end{array}
\end{equation}
The derivatives are
$$\begin{array}{ll}
\displaystyle\frac{\partial H}{\partial{\cal
E}}=\!\!&\!-\displaystyle\frac{H}{2\cal E}-\frac{1}{\kappa\sqrt
E}\left[(K_xE^x+K_yE^y){\cal
A}-\frac{1}{2}\left(\displaystyle\frac{E^x{E^y}'-{E^x}'E^y}{E^xE^y}\right)^2
{\cal
E}\right.\\[5mm]
&+\left.\displaystyle\frac{1}{2}\frac{(E^xE^y)'}{E^xE^y}\,{\cal
E}'-{\cal E}''\right],\\[8mm]
\displaystyle\frac{\partial H}{\partial{\cal
E}'}=&\!\displaystyle\frac{1}{\kappa\sqrt{E}}\left[{\cal
E}'-\frac{(E^xE^y)'}{E^xE^y}\,{\cal E}\right],
\hspace{5mm}\mbox{and}\hspace{5mm}\displaystyle\frac{\partial
H}{\partial{\cal E}''}=\frac{1}{\kappa\sqrt{E}}\,\cal E.
\end{array}$$
With the new constraints $\cal A$ and $B$ taken into account, we
have the following weak equivalences
\begin{equation}
\frac{\partial H}{\partial{\cal
E}}\approx\frac{E^x{E^y}''+{E^x}''E^y}{\kappa\, E^xE^y},
\hspace{5mm} \frac{\partial H}{\partial{\cal E}'}\approx0,
\hspace{5mm} \frac{\partial H}{\partial{\cal
E}''}\approx\frac{1}{\kappa}.
\end{equation}
Inserting this into (\ref{AH}) gives
\begin{equation}
\{{\cal
A},H[N]\}\approx-\frac{N}{E^xE^y}\,({E^x}''E^y+E^x{E^y}'')-N''.
\end{equation}
So we see that, as in the case of  $B$ with diffeomorphisms, $\cal
A$ is not invariant under the local action of $H(z)$, so the full
local Hamiltonian constraint becomes second-class, like the local
diffeomorphism constraint $C(z)$. If we choose a lapse function $N$
linear in $z$ and introduce the further constraint
\begin{equation}\label{D}
D:={E^x}''E^y+E^x{E^y}''
\end{equation}
then the constraint ${\cal A}$ is preserved under evolution.


For reasons that become more clear in later calculations, we make the more specialized choice $N=N(t)$. With the additional constraint $\partial_z N=0$ on the Lagrange multiplier $N$ the system remains consistent, as can be checked using the constraint algebra. With this assumption the Hamiltonian constraint is reduced to a global condition $H[N]=0$; the associated symmetry transformation is an evolution  in a global time. This choice is in accordance with the form of the metric (\ref{metric}), where the choice $N=1$ is even more special \footnote{The choice $N=N(t)$ is also essential for calculations of the algebra of constraints including  equation (\ref{HD}) and $\{U_\rho,H[N]\} \approx 0$. The constraints $U_\rho$ are  defined in equation (\ref{Udef}).}.

The constraint $D$ is interesting. Imposing $B=0$, we may express
\begin{equation}\label{Lb}
E^x=Le^{-\beta}\hspace{5mm}\mbox{and}\hspace{5mm}E^y=Le^\beta
\end{equation}
and, after insertion into (\ref{D}), the constraint equation $D=0$
becomes $2L(\partial_z^2L +(\partial_z \beta)^2 L) = 0$; this is the
spatial part of the Einstein equation (\ref{1stein}).

Now $D$ Poisson-commutes trivially with $\cal A$ and $B$. Its
Poisson bracket with $C[n]$, given $n'=0$, is
\begin{equation}
\{D[f],C[n]\}=-nD[f']
\end{equation}
and vanishes weakly. However, the analysis is not complete since we
have to be sure that $D=0$ is preserved under the Hamiltonian
constraint.

\subsection{Consistency of the reduced system}
\label{consistency}

Taking into account the constraints $\cal A$ and $B$, and under the condition $N=N(t)$, the Poisson bracket of $D$ with $H$ is, after integration by parts,
\begin{equation}
\label{HD}
\{D[f],H[N]\}\approx \displaystyle\int{\rm
d}z\,f\left[
N(K_x{E^x}''+K_y{E^y}''+K_x''E^x+K_y''E^y)\right].
\end{equation}
So far we have reduced a system on the six dimensional phase
space with two first-class local constraints, $H(z)$ and $C(z)$, to
one with five second-class local constraints $H(z)$, $C(z)$, ${\cal
A}(z)$, $B(z)$ and $D(z)$ and two global evolution generators
$H[N(t)]$ and $C[n(t)]$. Numerically five constraints per space
point would suffice to reduce six phase-space functions to one,
corresponding to the free function $\beta$ in (\ref{1stein}), but consistency under
time evolution requires more.

Even with the assumption that $N$ is independent of $z$, $D$ does
not weakly Poisson-commute with $H[N]$. The bracket is equivalent to
\begin{equation}\label{dh}
-N\int{\rm d}z\,f(K_x''E^x+K_y''E^y+K_x{E^x}''+K_y{E^y}'')
=:-N\int{\rm d}z\,f(z)J(z).
\end{equation}
The new constraint $J(z)$ can be expressed as a sum of similar terms including the derivative of $D$,
\begin{equation}\label{FU}
J=E^x(K_x-{E^y}')''+E^y(K_y-{E^x}')''+{E^x}''(K_x-{E^y}')+{E^y}''(K_y-{E^x}'
)+D',
\end{equation}
or, alternatively,
\begin{equation}\label{FV}
J=E^x(K_x+{E^y}')''+E^y(K_y+{E^x}')''+{E^x}''(K_x+{E^y}')+{E^y}''(K_y+{E^x}'
)-D'.
\end{equation}
The Poisson bracket of $J$ with $H$ is weakly equal to the second
derivative of $D$ (using $\cal A$ and $B$ in the equivalence) plus
additional terms
\begin{equation}\label{FH}
\begin{array}{l}
\{J[f],H[N]\}\approx\;N\displaystyle\int{\rm
d}z\,f\left\{D''-2\left(\displaystyle\frac{{E^x}'}{E^x}+
\displaystyle\frac{{E^y}'}{E^y}\right)({E^x}'{E^y}'-K_xK_y)'\right.\\[4mm]
\left.+2\left[\left(\displaystyle\frac{{E^x}'}{E^x}\right)^2+
\left(\displaystyle\frac{{E^y}'}{E^y}\right)^2\right]\left({E^x}'{E^y}'-K_xK
_y\right)
+4({E^x}''{E^y}''-K_x'K_y')\right\}.
\end{array}
\end{equation}
We clearly need to check the Poisson
bracket of $\{J,H\}$ with $H$. The constraints descended from $J$
contain higher and higher derivatives so this leads to an infinite
tower of constraints; the system in this form is inconsistent. On
the other hand, we know from the metric formulation of (\ref{metric}) and (\ref{1stein}) that
there is a consistent formulation for non-colliding waves with one
configuration degree of freedom per point in light-cone coordinates.
Obviously, there must be relations between the constraints to reduce
the number of independent ones.

This observation suggests an obvious solution to the apparent
inconsistency.
We can restrict the phase space
variables at the kinematical level so that they only support left- or
right-moving waves \footnote{There is precedent in the literature for this.
Scalar fields in 1+1 dimensions with the same property are called
``self-dual fields" and are quantized, for example, by Floreanini and
Jackiw \cite{JF}. The restriction of gravitational field variables to either left or
right moving waves will lead to analogous bracket relations as those
for the self-dual fields with corrections in the part of space-time
where the wave factor $\beta$ is non-constant.}.
The constraint $J$ in the form (\ref{FU}) or (\ref{FV}) weakly
vanishes when $K_x=\pm{E^y}'$ and $K_y=\pm{E^x}'$. Then $J$ and
$\{J,H\}$ are essentially $D'$ and $D''$ and so also weakly vanish.
Hence we impose either the ``right-moving"
\begin{equation}
\label{Udef}
U_x:=K_x-{E^y}'=0 \hspace{5mm}\mbox{and}\hspace{5mm}
U_y:=K_y-{E^x}'=0
\end{equation}
or the ``left-moving"
\begin{equation}
V_x:=K_x+{E^y}'=0 \hspace{5mm}\mbox{and}\hspace{5mm}
V_y:=K_y+{E^x}'=0
\end{equation}
as primary constraints. As shown in Appendix A these relations, together with the equation of motion, provide a consistent solution to the Einstein equations in terms of the triad and canonical momenta. The relations also anticipate that the canonical momentum of the metric variable $E^\rho$ is equal to $\pm$ its spatial derivative for pp-waves.

In the following we work with the right-moving constraints $U_\rho$.
The Poisson brackets of $U_x$ and $U_y$ are
\begin{equation}
\label{UU}
\begin{array}{l}
\{U_x(z),U_y(z')\}=\{U_y(z),U_x(z')\}=2\kappa\delta'(z-z')=
2\kappa\displaystyle\frac{\partial}{\partial z}\,\delta(z-z')\\[2mm]
=-\{U_y(z'),U_x(z)\}.
\end{array}
\end{equation}
Note the antisymmetry in $z$ and $z'$ in spite of the symmetry under
the exchange of $U_x$ and $U_y$!

These right moving constraints have non-vanishing Poisson brackets with $B$
\begin{equation}
\{U_x(z),B(z')\}=\kappa E^y(z)\delta(z-z'), \text{  and  }
\{U_y(z),B(z')\}=\kappa E^x(z)\delta(z-z').
\end{equation}
Introducing the multipliers $u_\rho$ (not to be confused with the
light-cone coordinate $u$ in the introduction) for the constraints
$U_\rho$ and $h$ for $D$ we have the Poisson bracket
\begin{equation}
\{U_x[u_x],D[h]\}=\kappa\displaystyle\int{\rm
d}z\left\{ [u_x(z)h(z)]''+u_x(z)''h(z) \right\} E^y(z),
\end{equation}
and a similar relation for $\{U_y[u_y],D[h]\}$. The Poisson brackets
$\{ U_\rho, H[N(t)] \}$
and $\{ U_\rho, C[n(t)] \}$ vanish weakly. Thus, the constraints
$U_\rho$ are compatible with time evolution and their introduction
solves the problem of the infinitely many constraints, thus making
time evolution consistent. On the other hand, this introduction
increases the number of second-class constraints to seven, which is
definitely too many. What remains to solve is this apparent
overconstraining of the system.

Physically the reason for the constraints $U$ or $V$ lies in the
fact that the full Hamiltonian constraint of plane gravitational
waves applies to modes going both into the positive and the negative
$z$ direction and their mutual interaction. A superposition of left-
and right- moving waves would introduce complicated interactions and
spoil the simple form of the metric. In section \ref{evolution} we
will see that under the conditions $U_\rho=0$ or
$V_\rho=0$ the Hamiltonian constraint generates simple plane wave
propagation.

\section{Dirac brackets}
\label{DB}

In this section we construct the Dirac brackets of the local
second-class constraints step by step (see below), according to
algebraic relationships. The algebraic Poisson bracket structure
associates the second-class constraints into two ``pairs", $({\cal
A},B)$ and $(U_x,U_y)$, and three single constraints $D$, $H$, and
$C$. In addition $\cal A$ and $B$ do not contain derivatives and so
are actually associated to each point $z$ separately. In the course
of the analysis the constraints $C$ and $H$ turn out to be
dependent, more precisely, equivalent to $D$, so that the set of
independent constraints reduces to the convenient number of five. An
odd number of second-class constraints (per space point $z$) may
appear incompatible with the standard construction of Dirac brackets
\cite{HT}, but not all of them are related exactly to one point,
some of them contain derivatives.

For a mechanical system with second-class constraints $C_i$,
$i=1,2,\ldots,2n$ the Dirac bracket of two phase space functions $F$
and $G$ is defined as
\begin{equation}\label{Dirac}
\{F,G\}_{\rm D}=\{F,G\}-\{F,C_i\}M_{ik}^{-1}\{C_k,G\}
\end{equation}
in terms of Poisson brackets. The matrix $M_{ik}^{-1}$ is the
inverse of the matrix $M_{ik}=\{C_i,C_k\}$ of the Poisson brackets
among the constraints. After the Dirac brackets are constructed, the
constraints can be imposed strongly. This reduces the system to its
physical degrees of freedom.

A helpful fact about Dirac brackets is that they can be constructed
recursively, i.~e. the construction of equation (\ref{Dirac}) can be
carried out for any subset of second-class constraints, provided the
matrix of their Poisson brackets is invertible \cite{HT}. After
imposing these constraints strongly the procedure can be repeated
with the preliminary Dirac brackets replacing the Poisson brackets
in equation (\ref{Dirac}). This possibility greatly facilitates the
work with our constraints. In field theory, of course, the matrix
multiplication in (\ref{Dirac}) implies integration.

\subsection{Dirac brackets, version D1}
Beginning with the pair $({\cal A},B)$ we have the Poisson brackets
(\ref{pair}) and
\begin{equation}\label{M}
M_{ik}^{-1}(z,z')=\frac{1}{\kappa}\,\delta(z-z')\left(\begin{array}{rc}0&1\\
-1&0\end{array}
\right).\end{equation} The ensuing Dirac brackets, version D1, are
explicitly
\begin{equation}
\begin{array}{l}
\{F(z),G(z')\}_{\rm D1}=\{F(z),G(z')\}-
\displaystyle\frac{1}{\kappa}\int{\rm d}z''\{F(z),{\cal
A}(z'')\}\{B(z''),G(z')\}\\[3mm]+\displaystyle\frac{1}{\kappa}\int{\rm
d}z''\{F(z),B(z'')\}\{{\cal A}(z''),G(z')\}.\end{array}
\end{equation}
Due to the appearance of $\cal A$ in both the integrals on the right
hand side, neither the Dirac brackets of the variables $E^\rho$,
$K_\rho$, nor those of the remaining constraints differ from the
corresponding Poisson brackets. We can simply impose $\cal A$ and
$B$ strongly. When this is done, $U_x$, $U_y$, and $D$ are
untouched, whereas $C$ and $H$ are simplified considerably: The
diffeomorphism constraint drops its term $-\frac{1}{\kappa}{\cal
E}'{\cal A}$ and becomes
\begin{equation}
C=\frac{1}{\kappa}\left(K_x'E^x+K_y'E^y\right),
\end{equation}
whereas the Hamiltonian constraint boils down to
\begin{equation}
H=-\frac{1}{\kappa}\left[K_xK_y+{E^x}'{E^y}'-(E^xE^y)''\right)]
\end{equation}
with the second-derivative term not contributing to integrals with a
$z$-independent test function. Without this term the last expression
for $H$ is similar to the Hamiltonian of two free Klein-Gordon
fields, the non-linearity of GR is now hidden in $D$, which is not
conserved under the evolution generated by the Hamiltonian
constraint. In the simplest case of constant lapse and shift,
e.\,g. $N=n=1$, $D$ commutes weakly with $C$,
\begin{equation}
\{D,C[1]\}=E^x{E^y}'''+{E^x}'{E^y}''+{E^x}''{E^y}'+{E^x}'''E^y=D',
\end{equation}
but not with $H$,
\begin{equation}
\{D,H[1]\}=E^xK_x''+{E^x}''K_x+E^yK_y''+{E^y}''K_y, \end{equation}
as long as we do not introduce the constraints $U_\rho$.

\subsection{Dirac brackets, version D2}
The next pair of second-class constraints, $(U_x,U_y)$, has the
mutual Poisson brackets (\ref{UU}). To construct the inverse of the
matrix of these brackets we need the inverse of the derivative of a
$\delta$-function, denoted by $\delta^{(-1)}$, which satisfies the
relation
\begin{equation}
\int{\rm d}z''\,\delta'(z-z'')\delta^{(-1)}(z''-z')=\delta(z-z').
\end{equation}
Obviously $\delta^{(-1)}(z-z')$ is a step function plus an additive
constant that is adjusted by demanding antisymmetry
\footnote{Up to boundary terms that do not apply when $z$ is in the interval $(z_1,z_2)$.}
\begin{equation}\label{xy}
\delta^{(-1)}(z-z')=\frac{1}{2}\,{\rm sign}(z-z')\;.
\end{equation}
We construct the matrix $N_{ik}^{-1}$ that plays an analogous role as $M_{ik}^{-1}$ in (\ref{M}),
\begin{equation}
N_{ik}^{-1}(z,z')=\frac{1}{4\kappa}\,{\rm
sign}(z-z')\left(\begin{array}{cc}0&1\\1&0
\end{array}\right).
\end{equation}
With this matrix the next version of Dirac brackets\\ $ \{F,G\}_{\rm
D2}=\{F,G \}-\{F,U_x\}N_{ik}^{-1}\{U_y,G\}, $ becomes 
\begin{equation}\label{FG}
\begin{array}{l}
\{F(z),G(z')\}_{\rm D2}=\{F(z),G(z')\}\\[4mm]
-\displaystyle\frac{1}{4\kappa}\int_{z_-}^{z_+}{\rm
d}z''{\rm d}z'''\{F(z),U_x(z'')\}\,{\rm
sign}(z''-z''')\{U_y(z'''),G(z')\}\\[4mm]
-\displaystyle\frac{1}{4\kappa}\int_{z_-}^{z_+}{\rm d}z''{\rm
d}z'''\{F(z),U_y(z'')\}\,{\rm sign}(z''-z''')\{U_x(z'''),G(z')\}.
\end{array}
\end{equation}
(The D1 brackets are the same as the Poisson brackets so the label is omitted.) 
In particular, the Dirac brackets of the remaining fundamental variables are the following
\begin{equation}
\begin{array}{l}
\{K_x(z),K_x(z')\}_{\rm D2}=\{K_y(z),K_y(z')\}_{\rm D2}=0,\\[3mm]
\{K_x(z),K_y(z')\}_{\rm
D2}=-\displaystyle\frac{\kappa}{2}\,\delta'(z-z'),\\[3mm]
\{K_x(z),E^x(z')\}_{\rm D2}=\{K_y(z),E^y(z')\}_{\rm
D2}=\displaystyle\frac{\kappa}{2}\,\delta(z-z'),\\[3mm]
\{K_x(z),E^y(z')\}_{\rm D2}=\{K_y(z),E^x(z')\}_{\rm D2}=0,\\[3mm]
\{E^x(z),E^x(z')\}_{\rm D2}=\{E^y(z),E^y(z')\}_{\rm D2}=0,\\[3mm]
\{E^x(z),E^y(z')\}_{\rm D2}=\displaystyle\frac{\kappa}{4}\:{\rm
sign}(z-z').
\end{array}
\end{equation}
The bracket relations between $E^x$ and $E^y$ may look awkward due
to non-locality.  This is explained by the form of the constraints
$U_x$ and $U_y$. Integrating them yields the $E$'s in form of an
integral over $K$.  The expression
$$E^y(z)=\frac{1}{2}\left[\int_{z_-}^zK_x(z'){\rm
d}z'-\int_z^{z_+}K_x(z'){\rm
d}z'\right]=\frac{1}{2}\int_{z_-}^{z_+}{\rm sign}(z-z')\,K_x(z'){\rm
d}z'$$ and its counterpart $E^x$ from $U_y=0$ make the non-locality
of their Dirac brackets plausible. These brackets are of the same
form as those of the self-dual fields of  \cite{JF}, see section V.

To impose the $U$ constraints strongly, we can express the $K$'s in
terms of the $E$'s or vice versa, or choose one of the canonical
pairs $(K_x,E^x)$ and $(K_y,E^y)$ as fundamental variables. To
preserve the canonical structure, the latter choices would seem to be
preferable, but
in different calculations different choices may be suitable.

After the $U$'s are imposed strongly, $C(z)$ and $H(z)$ become
equivalent to $\frac{1}{\kappa}D(z)$, explicitly
\begin{equation}
C=\frac{1}{\kappa}\,D+\frac{1}{\kappa}\,(U_x'E^x+U_y'E^y)
\end{equation}
and
\begin{equation}
H=\frac{1}{\kappa}\,D-\frac{1}{\kappa}\,(U_xU_y+U_x{E^x}'+U_y{E^y}'),
\end{equation}
this means that finally the number of independent local constraints
is reduced to five and that $D(z)$ now implies also the global
constraints $H[N]$ and $C[n]$. So the constraints $U_\rho$
themselves solve the problem of overconstraining that arose after
their introduction. Further, the fact that $U_\rho$ ($V_\rho$) lead
to $H[1]=\pm D[1]$ confirms that $U/V$ single out left/right moving
wave modes. The integrated Hamiltonian constraint with $N\equiv1$
becomes
\begin{equation}
H[1]=-\frac{2}{\kappa}\int{\rm
d}z\,{E^x}'(z)\,{E^y}'(z)=-\frac{2}{\kappa}\int{\rm
d}z\,K_x(z)\,K_y(z).
\end{equation}
Finally $D$ commutes with the total Hamiltonian, which is now (for
$N=1$ and $n=0$, according to the assumption in (\ref{metric})) just
$H[1]$,
\begin{equation}\label{HD2}
\{D(z),H[1]\}_{\rm D2}=D'(z)\approx0.
\end{equation}

\subsection{The final Dirac brackets}
At this point we have one phase space degree of freedom, represented
equivalently by one of the above-mentioned pairs of variables, one
local constraint $D(z)$ per point $z$ and one global one, $H[1]$,
which is at the same time the generator of time evolution. The
constraints $D(z)$ are second-class and their Dirac brackets,
version D2, are rather complicated.
\begin{equation}
\{D(z),D(z')\}_{\rm
D2}=\kappa\left[f(z,z')\delta'''(z-z')+g(z,z')\delta'(z-z')+h(z,z')\delta^{(
-1)}(z-z')\right],
\end{equation}
where $\delta^{(-1)}(z-z')$ was introduced in (\ref{xy}) and
\begin{equation}
\begin{split}
f(z,z') = \frac{1}{2}\left[ E^x(z)E^y(z')+E^x(z')E^y(z) \right],\\
g(z,z') = \frac{1}{2}\left[ E^x(z){E^y}''(z')+
{E^x}''(z)E^y(z')\right.\\
\left. + E^x(z'){E^y}''(z)+{E^x}''(z')E^y(z)\right],\\
h(z,z') = \frac{1}{2}\left[
{E^x}''(z){E^y}''(z')+{E^x}''(z'){E^y}''(z) \right].
\end{split}
\end{equation}
Let's denote by $\Delta(z,z')$ the inverse of $\{D(z),D(z')\}_{\rm
D2}$, needed in the construction of the final Dirac brackets
\begin{eqnarray}\label{D2}
&&\{F(z),G(z')\}_{\rm D}=\{F(z),G(z')\}_{\rm D2}-\\[2mm]
&&(\kappa)^{-1}\int{\rm d}z''\,{\rm d}z'''\{F(z),D(z'')\}_{\rm
D2}\,\Delta(z'',z''')\,\{D(z'''),G(z')\}_{\rm D2}.\nonumber
\end{eqnarray}
It is an antisymmetric function satisfying
\begin{equation}\label{inv}
\int{\rm d}z''\,\{D(z),D(z'')\}_{\rm
D2}\,\Delta(z'',z')=\delta(z-z').
\end{equation}
We do not have the full solution to this equation. In Appendix
\ref{IDD} we calculate an approximation, demonstrating some
qualitative features of the canonical structure rather than giving
the exact Dirac brackets. In the following $\Delta$ is understood as
this approximation and $\{\;,\;\}_{\rm D}$ as a representative part
of the full Dirac bracket, constructed with $\Delta$.

As already mentioned, when we apply five local constraints strongly,
there remains one free variable. If we choose $E^x$, our fundamental
Dirac brackets are those of $E^x$ at different points, constructed
according to (\ref{D2}). For this purpose we need the bracket
\begin{eqnarray}\label{ED}
\{E^x(z),D(z'')\}_{\rm D2}&=&E^x(z'')\,\{E^x(z),{E^y}''(z'')\}_{\rm
D2}+{E^x}''\,(z'')\{E^x(z),E^y(z'')\}_{\rm D2}\nonumber\\
&=&\frac{\kappa}{2}\,E^x(z'')\,\delta'(z-z'')+\frac{\kappa}{4}\,{E^x}''(z'')\,
{\rm sign}(z-z'');
\end{eqnarray}
$\{D(z'''),E^x(z')\}_{\rm D2}$ is calculated analogously. For our
approximation of $\{E^x(z),E^x(z')\}_{\rm D}$ we take only the
$\delta'$ parts of these brackets. In the following $\Delta$,
calculated in (\ref{Delta}), is more conveniently expressed in terms
of anti-derivatives of the $\delta$ functions,
$$\delta^{(-3)}(z-z')=\frac{1}{4}\,|z-z'|(z-z'),\hspace{5mm}
\delta^{(-5)}(z-z')=\frac{1}{48}\,|z-z'|(z-z')^3.$$ Putting these
ingredients together, we have
\begin{eqnarray*}
&&\{E^x(z),E^x(z')\}_{\rm D}\approx\frac{\kappa}{4}\int{\rm
d}z''\,{\rm
d}z'''\,E^x(z'')\,\delta'(z-z'')\left[\frac{\delta^{(-3)}(z''-z''')}
{\Lambda\left(\frac{z''+z'''}{2}\right)}+\right.\\[3mm]
&&\left.\frac{3}{4}\frac{\left(\Lambda'\left(\frac{z''+z'''}{2}\right)\right
)^2}
{\Lambda^3\left(\frac{z''+z'''}{2}\right)}\,\delta^{(-5)}(z''-z''')\right]E^
x(z''')
\,\delta'(z'''-z').
\end{eqnarray*}
After integrating the $\delta'$ functions by parts we expand
$E^x(z)$ and its first derivative around $\bar z=(z+z')/2$
$$E^x(z)\approx
E^x(\bar z)+{E^x}'(\bar z)\,\frac{z-z'}{2}+\ldots$$ and $E^x(z')$,
and make use of
$$\delta^{(-1)}(z-z')\cdot(z-z')^2=\delta^{(-2)}(z-z')\cdot(z-z')=2\,\delta^
{(-3)}(z-z').$$
With all the variables evaluated at $\bar z$ (so that $\Lambda$
corresponds to $\Lambda_0$ in Appendix B) and inserting finally
$\Lambda=L^2$ and $E^x=Le^{-\beta}$, we find
\begin{equation}
\label{EE} \{E^x(z),E^x(z')\}_{\rm D}\approx
\displaystyle\frac{\kappa}{8}\,e^{-2\beta}\,{\rm
sign}(z-z')\left[1-\left(2(\beta')^2+\frac{5}{4}\,\beta''\right)(z-z')^2\right].
\end{equation}
In the flat space-time regions $z_1<z<z_-$ and $z_+<z<z_2$ of our
coordinate domain, where $\beta=0$ (and $L''=0$) the field $E^x$
satisfies bracket relations analogous to the commutation relations
of self-dual Klein-Gordon fields, considered
in \cite{JF}, which are constructed by restriction to waves
going into one direction. The correction in the brackets for $E^x$
is expressed purely in terms of the wave factor $\beta$. Although
the approximation is rather qualitative, it is quite instructive for
some insight into the influence of the gravitational Hamiltonian in
the canonical structure of the self-dual fields \cite{JF}.

For $\beta\neq0$ the above expression can be interpreted as a
low-order approximation of a gravitational correction. Were this
quantized, this would appear as a variable Planck constant, as
suggested by Hossenfelder \cite{SH}, or a variable gravitational
constant. Other corrections, however, do not fit into this scheme,
they give rise to qualitatively different terms from (\ref{EE}).

\subsection{Time evolution}
\label{evolution}

The time evolution of a phase space function $F$ is generated by its
Dirac bracket with the total Hamiltonian. As already stated in the
preceding section, by virtue of the $U$'s $H[N]$ becomes equivalent
to $C[N]$ and so the total Hamiltonian becomes a generator of a
rigid translation. (Had we chosen the $V$ constraints, $H[N]$ would
be equivalent to $-C[N]$.) This equivalence allows to introduce
$C[N]$ as a true total Hamiltonian, when we choose a lapse function.
The most convenient choice $N\equiv1$ means a constant unit of time.

The Hamiltonian constraint $H_{\rm tot}=H[1]=C[1]$ being first-class
at every stage, its Dirac
brackets with any phase space function are equal to the corresponding
Poisson
brackets, which are equal to the $z$-derivative, according to the
nature of $C[1]$ as translation generator. Hence,
\begin{equation}
\dot{F}(x)=\{F(z),H[1]\}_{\rm D}=\{F(z),C[1]\}=F'(z).
\end{equation}
Equivalence of $H[N]$ with $\pm C[N]$ simply means that time
evolution is a rigid space translation to the left or to the right,
the same relation that characterizes self-dual fields \cite{JF}.

This completes the Einstein equation by making all the variables
depend on $t-z$ (or $t+z$, alternatively).  So we have recovered the
classical equation of motion (\ref{1stein}) in a much reduced phase
space.  One can describe the system with a single function, e.~g.
$E^x(z)$, on $(z_1,z_2)$ subject to the second-class constraint $D$.

\section{The Immirzi parameter and the polarization angle}
\label{orthog}

In this section we return to the orthogonality of the connection
components (\ref{aort}) to show that this is satisfied by the
reduced model. After strong imposition of all the constraints the
two-vectors $\vec A_x=({A_x}^1,{A_x}^2)$ and $\vec
A_y=({A_y}^1,{A_y}^2)$ are orthogonal and there arises a simple
relation between the angle $\alpha$ between
$\vec{E^x}=(E^x_1,E^x_2)$ and $\vec A_x=(A_x^1,A_x^2)$ and the
corresponding angle $\bar\alpha$.

The variables in polar coordinates from \cite{BD1}, with $\beta=\bar\beta=0$, corresponding to the gauge $\xi=\eta=0$ are
\begin{equation}\label{pol}
\begin{array}{l}
E^x_1=E^x, \hspace{4mm} E^x_2=0, \hspace{4mm} E^y_1=0, \hspace{4mm}
E^y_2=E^y;\\[4mm]
A_x^1=A_x\cos\alpha, \hspace{2mm} A_x^2=A_x\sin\alpha, \hspace{2mm}
A_y^1=-A_y\sin\bar\alpha, \hspace{2mm} A_y^2=A_y\cos\bar\alpha.
\end{array}
\end{equation}
From elementary calculations of the connection components $\Gamma$
in terms of the $E$'s we find in the gauge $\beta=\bar\beta=0$
\begin{equation}
\Gamma_x^1=\Gamma_y^2=0, \hspace{5mm} \Gamma_x^2=-{E^y}',
\hspace{3mm} \Gamma_y^1=-{E^x}',
\end{equation}
so that the diagonal components $A_x^1$ and $A_y^2$ contain only
extrinsic curvature,
\begin{equation}\label{K}
A_x^1=A_x\cos\alpha=\gamma K_x^1, \hspace{5mm}
A_y^2=A_y\cos\bar\alpha=\gamma K_y^2.
\end{equation}
On the other hand, from the Gau\ss\ and the polarization constraint
$\{\xi,H\}=0$ we obtain (\cite{BD1}, (A.14))
\begin{equation}
A_x^2=A_x\sin\alpha=\Gamma_x, \hspace{5mm}
A_y^1=-A_y\sin\bar\alpha=-\Gamma_y,
\end{equation}
thus the off-diagonal components are purely composed from
$\Gamma$'s. Now the vectors $\vec A$ have acquired the form
\begin{equation}
{\vec A}_x=(\gamma K_x,-{E^y}'), \hspace{5mm} {\vec
A}_y=(-{E^x}',\gamma K_y),
\end{equation}
where we have written $K_x=K_x^1$ and $K_y=K_y^2$. For the absolute
squares of these vectors we get
\begin{equation}
(A_x)^2=\gamma^2(K_x)^2+({E^y}')^2 \hspace{3mm}\mbox{ and
}\hspace{3mm} (A_y)^2=\gamma^2(K_y)^2+({E^x}')^2.
\end{equation}
With the constraints $U_\rho$ (or $V_\rho$) this becomes
\begin{equation}
A_x=K_x\sqrt{1+\gamma^2} \hspace{3mm}\mbox{ and }\hspace{3mm}
A_y=K_y\sqrt{1+\gamma^2}.
\end{equation}
Inserting into (\ref{pol}) and comparing with (\ref{K}),
$$A_x^1=K_x\sqrt{1+\gamma^2}\cos\alpha=\gamma K_x, \hspace{5mm}
A_y^2=K_y\sqrt{1+\gamma^2}\cos\bar\alpha=\gamma K_y,$$ leads to a
relation between the angles $\alpha$ and $\bar\alpha$ and the
Immirzi parameter:
\begin{equation}
\alpha=\bar\alpha={\rm arccot}\gamma,
\end{equation}
so in the end, after all gauge fixing, $\vec A_x$ and $\vec A_y$ are
orthogonal and orthogonality is compatible with the Hamiltonian
constraint, when the latter reduces to a translation generator.

\section{Summary and conclusion}
\label{sum}

The principal aim of our considerations is the loop quantization
of polarized gravitational plane waves to see whether dispersion
relations would be modified, or if there are other effects from the
granularity of the kinematic states such as a variable speed of
gravitation or a variable Planck constant. One way to handle a
quantum theory of plane waves is to quantize a more general system,
such as a model with plane symmetry, analogous to the Gowdy model
exploited here, and then to distinguish the subspace of left- or
right-going modes of the full Hilbert space. But, as the formulation
of basic operators in \cite{BD2} shows, this turns out to be quite
complicated.

In the present work we reduced the formalism of plane waves to the
physical degree of freedom at the classical level. To derive a
classical description of plane waves suitable for loop quantization we
started with the assumption of homogeneity of the spatial geometry
in the transversal directions and a coordinate system extending in
both directions beyond a gravitational pulse, so that the latter one
is embedded between two slabs of flat space in these coordinates.
The finite range of this coordinate system rendered the integrations
over the remaining spatial coordinate finite. This setting is
analogous to the Gowdy model described in \cite{BD1} and we used the
formalism developed therein. The reduction was completed with a
Dirac constraint analysis.

This was done as follows. The description starts with a symmetry
reduced model with three configuration space degrees of freedom and
the standard diffeomorphism and Hamiltonian constraints of GR,
giving one field degree of freedom per spatial point. The first step
of pp-wave reduction was carried out in section \ref{Breduction}. We
restricted the metric in the triad variables to the simple diagonal
form of (\ref{metric}) by introducing the constraint $B$.
Preservation of this constraint led us to introduce the secondary
constraints ${\cal A} = 0$ and $D=0$, the latter one being the
(spatial projection of) the classical Einstein equation \cite{MTW}.
Knowing that the symmetry reduced system cannot describe colliding
waves and requiring consistency in the sense that further secondary
constraints vanish, we introduced the right-  (or left-) moving
constraints $U_\rho$ (or $V_\rho$) in section \ref{consistency}.
After restricting the fields to left- or right-moving
modes by imposing constraints we find a system analogous to the
self-dual  fields -- also scalar fields propagating in one direction
-- described by Floreanini and Jackiw \cite{JF}.

We impose these constraints in a special coordinate system, fixed by
the constraint $B$, which breaks diffeomorphism invariance, but
makes the form of the constraints $U_\rho$ simple. Alternatively,
the constraints distinguishing left/right going modes can be
formulated in a fully diffeomorphism invariant manner. Work is underway on
this approach \cite{ppwave2}.


In the second step of the reduction we used the ``pair-wise"
structure of the Poisson algebra of the constraints to  recursively
construct Dirac brackets.  In section \ref{DB} using $\cal A $ and
$B$ and then $U_\rho$  we constructed the first two versions of
Dirac brackets. After imposing the constraints $U_\rho$, the
constraint $D(z)$ commutes with the Hamiltonian, but becomes
second-class by itself. At this point the evolution has become
trivial; all the complications are now in the D2-bracket relation of
$D(z)$ and $D(z')$. To complete the canonical treatment of the
problem, we constructed the final Dirac brackets. In the end the $D(z)$
are the remaining second-class constraints, leading to one
variable (we have chosen $E^x$) with a very non-trivial Dirac
bracket $\{E^x(z),E^x(z')\}_{\rm D}$, containing the step function
${\rm sign}(z-z')$, multiplied by an analytic function in $z-z'$. An
approximation is given in equation (\ref{EE}). It is clear from
this bracket that the canonical structure of the reduced system is
obscure when using this variable.

This makes a full quantization, i.~e. a formulation in terms of
operators and a Hilbert space, elusive -- we do not even know the
exact closed form of the Dirac brackets. The lowest-order terms,
however, turn out to be the analog of the commutation relation of a
linear self-dual field \cite{JF} plus gravitational corrections.

The result of our preliminary classical considerations gives no
suggestion of dispersion in these waves, which would provide an
indication of an energy-dependent speed of gravitation. The
reduction to left or right moving waves leads automatically to
the equivalence of the Hamiltonian to the generator of spatial
translations. This equivalence was not assumed from the beginning,
as in other approaches, for example in light-like coordinates, but
it appears as a result of the disentangling of otherwise colliding
modes by the constraints $U_\rho$. In this way this complete
reduction with the aid of Dirac brackets differs from other
canonical approaches such as \cite{AB}. In a quantum theory, based
on our classical analysis, where these generators are promoted to
operators, an analogous result can be expected.  The final Dirac
bracket does hint at a modification of the quantum relations.

The non-local bracket of $E^x(z)$ (\ref{EE}) suggests a modification
of the Planck constant (or the gravitational constant) in the first
approximation, rather than a variable speed of light. In the
framework of our approach (starting from unmodified GR) we expect
that the space-time texture arising in a quantum theory of gravity
would influence the fundamental structure of quantum theory, mainly
the commutators and the uncertainty relations derived from them.

Finally, a remark on the approach to the reduction: 
After specializing to a $z$-independent shift vector and a
lapse function of the same type in order to conserve $B$ and the
ensuing constraints we could have abandoned the local constraints
$C(z)$ and $H(z)$ and kept only the according global ones $C[1]$ and
$H[1]$. This would mean to start with a theory different from GR, a
theory without full diffeomorphism and time-reparametrization
invariance. Nevertheless, we would have arrived at the same results,
because on the constraint surface determined by $\cal A$, $B$, $U_x$ and $U_y$
the local constraints $H(z)$ and $C(z)$ are equivalent to $D(z)$.
For this reason we did not have to make explicit use of the local
diffeomorphism and Hamiltonian constraint in our work. 

\appendix

\section{The Ricci tensor for the metric ${\rm diag}\left(-1,
(E^y)^2, (E^x)^2, 1\right)$}

To introduce canonical variables, we calculate from the
Levi-Civit\`{a}
connection the extrinsic curvature components
$$K_x=\dot E^y, \hspace{2cm} K_y=\dot E^x.$$
as canonical conjugate variables to the metric variables $E^x$ and
$E^y$. In terms of these canonical variables we get the five
non-vanishing independent components of the Ricci tensor:
$$R_{00}=-\left(\frac{\dot K_y}{E^x}+\frac{\dot
K_x}{E^y}\right),\hspace{5mm}
R_{03}=-\left(\frac{K_y'}{E^x}+\frac{K_x'}{E^y}\right),$$
$$R_{33}=-\left(\frac{{E^x}''}{E^x}+\frac{{E^y}''}{E^y}\right),$$
$$R_{11}=E^y(\dot
K_x-{E^y}'')+\frac{E^y}{E^x}(K_xK_y-{E^x}'{E^y}'),$$
$$R_{22}=E^x(\dot
K_y-{E^x}'')+\frac{E^x}{E^y}(K_xK_y-{E^x}'{E^y}').$$ The vacuum
Einstein equations $R_{03}=0$ and $R_{33}=0$ are the constraints $C$
and $D$, divided by $-E^xE^y$, the remaining ones contain time
evolution. Here $C$ and $D$ appear as primary second-class
constraints, unless we smear out $C$ with a $z$-independent
function. As in the main text, all the constraints are consistent
when $U_\rho=0$ or $V_\rho=0$ and $\partial/\partial
t=\pm\partial/\partial z$. For a quantization,
also this approach requires Dirac brackets.\\

\section{The inverse of $\{D(z),D(z')\}_{\rm D2}$}
\label{IDD} In the defining equation (\ref{inv}) of $\Delta(z,z')$,
with (\ref{ED}) inserted, the derivative of the $\delta$ function is
shifted to its second argument: $\delta'(z-z'')=-\partial/\partial
z''\delta(z-z'')$, and analogously the third derivative. Then,
integrating by parts, we get
\begin{equation}\label{inv2}
\begin{array}{l}
\displaystyle\int{\rm
d}z''\left\{\frac{\partial^3}{\partial{z''}^3}\left[f(z,z'')\Delta(z'',z')\right]
+\displaystyle\frac{\partial}{\partial
z''}\left[g(z,z'')\Delta(z'',z')\right]\right\}\,\delta(z-z'')\\[2mm]
+\displaystyle\int{\rm
d}z''\,h(z,z'')\,\delta^{(-1)}(z-z'')\,\Delta(z'',z')=\delta(z-z').
\end{array}\end{equation}

Consider the third derivative of the first square bracket,
\begin{equation}
\left(\frac{\partial^3}{\partial{z''}^3}\,f\right)\Delta+3\,\left(\frac{\partial^2}
{\partial{z''}^2}\,f\right)\frac{\partial}{\partial
z''}\,\Delta+3\,\left(\frac{\partial}{\partial
z''}\,f\right)\frac{\partial^2}{\partial{z''}^2}\Delta+f\,\frac{\partial^3}
{\partial{z''}^3}\,\Delta.
\end{equation}
This is multiplied by $\delta(z-z'')$ in (\ref{inv2}), so we need
$f$ and its derivatives at $z''=z$.
\begin{eqnarray*}
f(z,z)=E^x(z)E^y(z), &&\left.\frac{\partial f}{\partial
z''}\right|_{z''=z}=\frac{1}{2}\left(E^x(z)E^y(z)\right)',\\[2mm]
\left.\frac{\partial^2f}{\partial{z''}^2}\right|_{z''=z}=\frac{1}{2}\,D(z)\approx0,
&&\left.\frac{\partial^3f}{\partial{z''}^3}\right|_{z''=z}=\frac{1}{4}\left(
3D'(z)- (E^xE^y)'''\right).
\end{eqnarray*}
Similarly $g(z,z)=D(z)$ and $\left.\partial/\partial
z''\,g(z,z'')\right|_{z''=z}=\frac{1}{2}\,D'(z)$, so the second
square bracket does not contribute anything, when $D\approx0$.
Denoting $E^x(z)E^y(z)=L^2(z)$ by $\Lambda(z)$, we may write
(\ref{inv2}) in the form
\begin{equation}\label{inv3}
\begin{array}{l}
\Lambda(z)\displaystyle\frac{\partial^3}{\partial
z^3}\,\Delta(z,z')+\frac{3}{2}\,\Lambda'(z)\frac{\partial^2}{\partial
z^2}\,\Delta(z,z')-\frac{1}{4}\,\Lambda'''(z)\Delta(z,z')\\[3mm]
+\displaystyle\int{\rm
d}z''\,h(z,z'')\delta^{(-1)}(z-z'')\,\Delta(z'',z')=\delta(z-z').
\end{array}\end{equation}
In this equation $\Delta$ can be considered as a Green function of
an integro-differential operator. Because of the required
antisymmetry of the Dirac bracket we are looking for an
antisymmetric Green function.

The dominant coefficient function is $\Lambda$, the square of the
background factor of the gravitational wave, its derivatives are
smaller. To find an approximative part of the solution of this
equation we
first look for a solution of the differential part and leave the
integral part for later iterative corrections. The leading term
contains a third derivative, therefore the leading term in $\Delta$
is expected to contain the
function
\begin{equation} \delta^{(-3)}(z-z')=\frac{1}{4}\,{\rm
sign}(z-z')\,(z-z')^2=\frac{1}{4}\,|z-z'|(z-z'),
\end{equation}
the third derivative of which is the delta function on the right
hand side. To find at least an approximation for $\Delta$, we make
an ansatz in the form of a product of this function by a symmetric
function of $z$ and $z'$. Further we assume this function to be
analytic in some neighborhood of $z=z'$. Written in terms of $z+z'$
and $z-z'$, this function has only even powers in $z-z'$, so that
\begin{equation}\label{an}
\Delta(z,z')=\frac{1}{4}\,{\rm
sign}(z-z')\left[a_2(z-z')^2+a_4(z-z')^4+\ldots\,\right]
\end{equation}
with the coefficients $a_i$ being (analytic) functions of $\bar
z:=\frac{z+z'}{2}$. For the derivatives with respect to $z$ we find
\begin{eqnarray*}
&&\Delta_{,z}(z,z')=\frac{{\rm
sign}(z-z')}{4}\left[2a_2(z-z')+\frac{1}{2}\,a_2'\,(z-z')^2+\ldots\;\right]\\[2mm]
&&\Delta_{,zz}(z,z')=\frac{{\rm
sign}(z-z')}{4}\left[2a_2+2a_2'(z-z')+
\left(\frac{a_2''}{4}+12a_4\right)(z-z')^2+\ldots\,\right]\\[2mm]
&&\Delta_{,zzz}(z,z')=a_2\delta(z-z')+\frac{{\rm
sign}(z-z')}{4}\left[3a_2'+\left(\frac{3a_2''}{2}+24a_4\right)(z-z')\right.\\
&&\hspace{2cm}+\left.\left(\frac{a_2'''}{8}+18a_4'\right)(z-z')^2+\ldots\;\right]
\end{eqnarray*}
Inserting this into the first line of (\ref{inv3}) gives
\begin{eqnarray}\label{inv4}
&&\Lambda a_2\delta(z-z')+\frac{3}{4}(\Lambda
a_2'+\Lambda'a_2)\,{\rm
sign}(z-z')\nonumber\\
&&+\left(\frac{3}{8}\,\Lambda
a_2''+\frac{3}{4}\,\Lambda'a_2'+6\Lambda a_4\right)|z-z'|
\\
&&+\left(\frac{1}{32}\,\Lambda
a_2'''+\frac{3}{32}\,\Lambda'a_2''-\frac{1}{16}\,\Lambda'''a_2+\frac{9}{2}\,
\Lambda
a_4'+\frac{9}{2}\,\Lambda'a_4\right)|z-z'|(z-z')+\ldots\nonumber
\end{eqnarray}
Note that $\Lambda$ and its derivatives are functions of $z$,
whereas the $a_i$ are functions of $\bar z$ and their derivatives
refer to this argument.

To express everything in terms of $\bar z$ and $z-z'$, we expand
$\Lambda(z)=\Lambda(\frac{z+z'}{2}+\frac{z-z'}{2})$ around $\bar z$:
\begin{equation}
\Lambda(z)=\Lambda_0+\frac{1}{2}\,\Lambda_0'(z-z')+\frac{1}{8}\,\Lambda_0''(
z-z')^2+\ldots
\end{equation}
where $\Lambda_0=\Lambda(\bar z)$. Inserting into (\ref{inv4}) and
rearranging terms gives
\begin{eqnarray*}
&&\Lambda_0a_2\delta(z-z')+\frac{3}{4}(\Lambda_0a_2'+\Lambda_0'a_2){\rm
sign}(z-z')+\\
&&\left(\frac{3}{8}\,\Lambda_0a_2''+\frac{9}{8}\,\Lambda_0'a_2'+\frac{3}{8}\
,\Lambda_0''a_2
+6\Lambda_0a_4\right)|z-z'|+\\
&&\left(\frac{1}{32}\,\Lambda_0a_2'''+\frac{9}{32}\,\Lambda_0'a_2''+\frac{15
}{32}\,\Lambda_0''
a_2'+\frac{1}{32}\,\Lambda_0'''a_2+\frac{9}{2}\,\Lambda_0a_4'+\frac{15}{2}\,
\Lambda_0'a_4\right)\times\\
&&|z-z'|(z-z')+\ldots
\end{eqnarray*}

The coefficient of the $\delta$ function on the right hand side of
(\ref{inv3}) is one, therefore $\Lambda_0a_2=1$. Now all primes
denote derivatives with respect to $\bar z$, so we can split off the
vanishing derivatives of $\Lambda_0a_2$ and have
\begin{eqnarray}\label{B8}
&&\delta(z-z')+\left(\frac{3}{8}\,\Lambda_0'a_2'+6\Lambda_0a_4\right)|z-z'|+
\\
&&\left(\frac{3}{8}\,\Lambda_0'a_2''+\frac{3}{4}\,\Lambda_0''a_2'+\frac{9}{2
}\,\Lambda_0a_4'
+\frac{15}{2}\,\Lambda_0'a_4\right)|z-z'|(z-z')+\ldots\nonumber
\end{eqnarray}
Inserting now $a_2=\Lambda_0^{-1}$ and setting the coefficient of
$|z-z'|$ equal to zero we find
\begin{equation}
a_4=\frac{1}{16}\frac{(\Lambda_0')^2}{\Lambda_0^3}
\end{equation}
or, in terms of $L$,
\begin{equation}
a_2=\frac{1}{L^2(\bar z)},\hspace{2cm}
a_4=\frac{1}{4}\frac{(L')^2(\bar z)}{L^4(\bar z)}.
\end{equation}
The coefficient of $|z-z'|(z-z')$ in the last term of (\ref{B8})
cancels, so our ansatz (\ref{an}) leads to an antisymmetric
approximation of the inversion of $\{D(z),D(z')\}_{\rm D2}$ in some
neighborhood of $z=z'$,
\begin{equation}\label{Delta}
\Delta(z,z')\approx\frac{|z-z'|(z-z')}{4L^2(\bar z)}\left[1+
\frac{1}{4}\frac{(L')^2(\bar z)}{L^2(\bar z)}\, (z-z')^2\right].
\end{equation}
This is an approximation to the differential part of the
integro-differential equation (\ref{inv3}), iterating the integral
part would not give contributions of this order, but only higher
anti-derivatives of $\delta$.

On the other hand, had we begun with the integral part, we would
have obtained $\delta'$ as first approximation, because the integral
contains the first anti-derivative. Beginning with this, we would
get higher and higher derivatives of the $\delta$ function in the
sequel. This does not show up in our $\Delta$ and so the ensuing
expression for the Dirac brackets reveals only part, although an
important part, of the consequences of imposing $D$ strongly.

\noindent{\large\bf Acknowledgement.} The work was supported by the
Ministry of Education of the Czech Republic, contract no. MSM
0021622409. We thank Hamilton College, Masaryk
University, and the Perimeter Institute for hospitality and support.
This research was supported in part by the Perimeter Institute for
Theoretical Physics. We are grateful to Klaus Bering for pointing out
reference \cite{JF} and for helpful discussions. We thank an 
anonymous referee for helping to clarify the canonical analysis.

\end{document}